# MODELING TRAFFIC MANAGEMENT IN ATM NETWORKS WITH OPNET

# Rohit Goyal[1], Raj Jain[1], Sonia Fahmy[1], Shobana Narayanaswamy[2]

1. **The Ohio State University, Department of Computer and Information Science, 2015 Neil Ave, DL 395, Columbus, OH 43210**
2. **MIL3 INC, 3400 International Drive, NW · Washington, DC 20008**

## ABSTRACT

Asynchronous transfer mode (ATM) is the new generation of computer and communication networks that are being deployed throughout the telecommunication industry as well as in campus backbones. ATM technology distinguishes itself from the previous networking protocols in that it has the latest traffic management technology and thus allows guaranteeing delay, throughput, and other performance measures. This in turn allows users to integrate voice, video, and data on the same network. Available bit rate (ABR) service in ATM has been designed to fairly distribute all unused capacity to data traffic and is specified in the ATM Forum's Traffic Management (TM4.0) standard. This paper will describe the OPNET models that have been developed for ATM and ABR design and analysis.

## INTRODUCTION

With the convergence of telecommunication and computer industries, computer networking is adopting a new paradigm called Asynchronous Transfer Mode (ATM). ATM was selected by the telecommunication (carrier) industry as the technology to deliver the Broadband Integrated Services Digital Network (B-ISDN) carrier service. ATM is designed to handle different kinds of communication traffic (voice, audio, video and data) in an integrated manner. It is first technology to promise seamless interworking between the LAN and WAN network environments. The international standards for ATM networks have been formulated by the ATM Forum and ITU-T.

### The ATM Quality of Service Model

ATM uses short, fixed size (53 bytes) cells with connection oriented data transfer. This facilitates packet based voice, video and data transmission in an integrated services network. The ATM Forum specifies that ATM connections belong to ATM service categories that support certain Quality of Service (QoS) requirements. The ATM-Forum Traffic Management Specification (TM4.0) defines five service categories for ATM networks. Each service category is defined using a traffic contract and a set of QoS parameters. The *traffic contract* is a set of parameters that specify the characteristics of the source traffic. The *QoS parameters* are negotiated by the source with the network, and are used to define the expected quality of service provided by the network. For each service category, the network guarantees the negotiated QoS parameters if the end system complies with the negotiated traffic contract. For non-compliant traffic, the network need not maintain the QoS objective.

Four traffic parameters are specified – Peak Cell Rate (PCR), Sustainable Cell Rate (SCR), Maximum Burst Size (MBS), and Minimum Cell Rate (MCR). A tolerance value for PCR, called Cell Delay Variation Tolerance (CDVT), is also specified. This defines the allowable variation from PCR for a connection's traffic to be conforming. The network determines conformance to PCR and SCR by using the Generic Cell Rate Algorithm (GCRA). The GCRA is a version of the token bucket algorithm that determines if an incoming cell is conforming to the traffic specifications. If a cell is non-conforming, the usage parameter control (UPC) function of the network may tag the cell by setting the CLP bit in the cell. The untagged and the aggregate cell streams are referred to as the CLP0 and CLP0+1 cell streams respectively.

The ATM Forum also defines six QoS parameters that specify the network performance objectives for the connection – Cell Loss Ratio (CLR), Cell Delay Variation (peak-to-peak CDV or 2-point CDV), Cell Transfer Delay (max CTD or mean CTD), Cell Error Ratio (CER), Severely Errored Cell Block Ratio (SECBR) and Cell Misinsertion Rate (CMR). Of these, the CDV, CTD and the CLR parameters are negotiated during connection setup, while the others are specified by the network. The CLR is further classified as being applicable for the CLP0 cell stream or the CLP0+1 cell stream.

Based on the QoS parameters, the ITU-T defines four QoS classes with defined values for some or all of the QoS parameters -- Class 1 (stringent class), Class 2 (tolerant class), Class 3 (bi-level class), and Class U (unspecified class). In addition, ATM switches are free to support additional QoS levels (or classes) for providing a finer granularity QoS support.

### ATM Service Categories

Based on the traffic parameters and the QoS parameters, the ATM forum specifies five service categories for the transport of ATM cells. The *Constant Bit Rate (CBR)* service category is defined for traffic that requires a constant amount of bandwidth, specified by a Peak Cell Rate (PCR), to be continuously available. The network guarantees that all cells emitted by the source that conform to this PCR will be transferred by the network with minimal cell loss, and within fixed bounds of maxCTD and peak-to-peak CDV.

The *real time Variable Bit Rate (VBR-rt)* class is characterized by PCR (and its associated tolerance), Sustained Cell Rate (SCR) and a Maximum Burst Size (MBS) in cells that controls the bursty nature of VBR traffic. The network attempts to deliver cells of conforming connections within fixed bounds of maxCTD and peak-to-peak CDV. *Non-real-time VBR* sources are also specified by PCR, SCR and MBS, but are less sensitive to delay and delay variation than the real time sources. The network does not specify any delay and delay variation parameters for the VBR-nrt service. Both VBR-rt and VBR-nrt are further divided into three categories based on whether CLR0 or CLR0+1 is specified as the CLR performance objective for the network, and whether tagging is allowed by the user or not.

The *Available Bit Rate (ABR)* service category is specified by a PCR and Minimum Cell Rate (MCR) which is guaranteed by the network. The bandwidth allocated by the network to an ABR connection may vary during the life of a connection, but may not be less than MCR. ABR connections use a rate-based closed-loop feedback-control mechanism for congestion control. The network tries to maintain a low Cell Loss Ratio by changing the allowed cell rates (ACR) at which a source can send. The *Unspecified Bit Rate (UBR)* class is intended for best effort applications, and this category does not support any service guarantees. UBR has no built in congestion control mechanisms. The UBR service manages congestion by efficient buffer management policies in the switch. A new service called Guaranteed Frame Rate (GFR) is being introduced at the ATM Forum and the ITU-T. GFR is based on UBR, but guarantees a minimum rate to connections. The service also recognizes AAL5 frames, and performs frame level dropping as opposed to cell level dropping.

### The OPNET Model

OPNET is a modeling and simulation tool [MIL31] that provides an environment for analysis of communication networks. The tool provides a three layer modeling hierarchy. The highest layer referred to as the network domain allows definition of network topologies. The second layer referred to as the node domain allows definition of node architectures (data flow within a node). The third layer (process domain) specifies logic or control flow among components in the form of a finite state machine.

### The OPNET ATM Model Suite

The OPNET ATM model suite (AMS) described in [MIL32] supports many of the characteristics of ATM networks. The model suite provides support for signaling, call setup and teardown, segmentation and reassembly of cells, cell transfer, traffic management and buffer management. Standard ATM nodes such as routers, stations, bridges, switches etc. are provided to facilitate building of common topologies used for the design and analysis of ATM networks.

Traffic management within AMS incorporates functions such as call admission control, policing using a continuous-state leaky bucket implementation (GCRA), call-based queuing, priority scheduling and collection of standard statistics such as end-to-end delay and end-to-end delay variation.

### Reference Topology

The example network topology used for the design and development of traffic management functions within AMS represents an N-source configuration shown in Figure 1. Source and destination end-systems are connected to a pair of ATM switches that communicate via a bottleneck link.

The node architecture for the end-system (source/destination) consists of AAL clients sending/receiving traffic to/from the AAL/ATM/PHY protocol stack. The AAL layer is responsible for segmentation of data traffic into AAL PDU's. The ATM layer (represented as four modules: management, layer, translation and switching) encapsulates the AAL PDU within the ATM cell and transmits the cell to the network. The management module is responsible for signaling. The translation module receives incoming traffic and directs it to the higher layer or back to the network based on the destination address.

Figure 1 The OPNET ATM Model

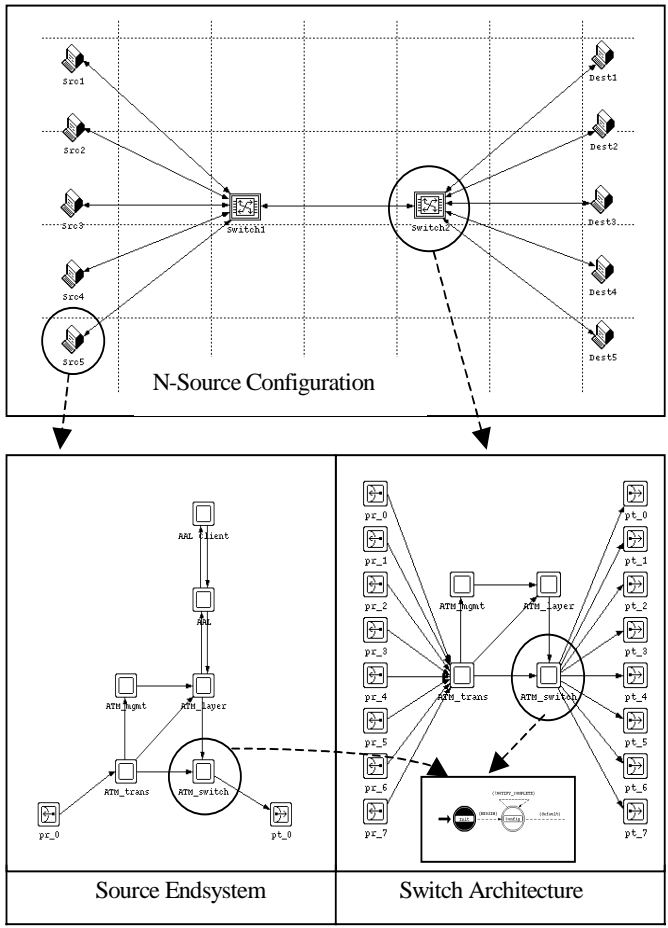

The node architecture for the switch consists of the ATM layer functions modeled as four module as described above. The switch can have many input and output ports.

ABR traffic management and feedback functions are implemented within the ATM switch module in the form of a finite state machine.

## Specification of QOS

AAL clients which represent sources of traffic specify their QoS requirement with the application traffic contract attribute. This requirement is a combination of service category, traffic parameters and QoS parameters for both incoming and outgoing directions that the source would like the network to provide. Traffic parameters include the PCR, MCR, SCR and MBS. QoS parameters include the CTD, CDV and CLR for both directions.

| Service Category | Requested Traffic Parameters | Requested QoS Parameters |
|---|---|---|

Application Traffic Contract

In order to be able to provide the requested QoS for a connection, intermediate devices may be configured to support various QoS levels. The switch buffer configuration attribute allows specification of QoS levels for each buffer. Cells streams belonging to different QoS levels may be buffered and serviced according to their QoS. The buffer configuration defines the buffer size, the maximum allocated bandwidth and minimum guaranteed bandwidth. The supported traffic parameters include PCR, MCR, SCR and MBS. The supported QoS parameters include CTD, CDV and CLR.

| Buffer Configuration | Supported Traffic Parameters | Supported QoS Parameters |
|---|---|---|

Switch Buffer Configuration

## ABR Traffic Management

ABR mechanisms allow the network to divide the available bandwidth fairly and efficiently among the active traffic sources. In the ABR traffic management framework, the *source end systems* limit their data transmission to rates allowed by the network. The network consists of switches that use their current load information to calculate the allowable rates for the sources. These rates are sent to the sources as feedback via *resource management (RM)* cells. The ABR traffic management model is a *rate-based end-to-end closed-loop* model.

There are three ways for switches to give feedback to the sources. First, each cell header contains a bit called Explicit Forward Congestion Indication (EFCI), which can be set by a congested switch. Such switches are called *binary* or *EFCI* switches. Second, RM cells have two bits in their payload, called the Congestion Indication (CI) bit and the No Increase (NI) bit, that can be set by congested switches. Switches that use only this mechanism are called relative rate marking switches. Third, the RM cells also have another field in their payload called explicit rate (ER) that can be reduced by congested switches to any desired value. Such switches are called Explicit Rate switches. RM cells are generated by the sources and travel along the data path to the *destination end systems*. The destinations simply return the RM cells to the sources.

Explicit rate switches normally wait for the arrival of an RM cell to give feedback to a source. However, under extreme congestion, they are allowed to generate an RM cell and send it immediately to the source. This optional mechanism is called backward explicit congestion notification (BECN).

Switches can use the virtual source/virtual destination (VS/VD) feature to segment the ABR control loop into smaller loops. In a VS/VD network, a switch can additionally behave both as a (virtual) destination end system and as a (virtual) source end system. As a destination end system, it turns around the RM cells to the sources from one segment. As a source end system, it generates RM cells for the next segment. This feature can allow feedback from nearby switches to reach sources faster, and allow hop-by-hop control.

At the time of connection setup, ABR sources negotiate several operating parameters with the network. The first among these is the peak cell rate (PCR). This is the maximum rate at which the source will be allowed to transmit on this virtual circuit (VC). The source can also request a minimum cell rate (MCR) which is the guaranteed minimum rate. The network has to reserve this bandwidth for the VC. During the data transmission stage, the rate at which a source is allowed to send at any particular instant is called the allowed cell rate (ACR). The ACR is dynamically changed between MCR and PCR. At the beginning of the connection, and after long idle intervals, ACR is set to initial cell rate (ICR). A complete list of parameters used in the ABR mechanism is given in [TM40].

Most resource management cells generated by the sources are counted as part of their network load in the sense that the total rate of data and RM cells should not exceed the ACR of the source. Such RM cells are called ''in-rate'' RM cells. Under exceptional circumstances, switches, destinations, or even sources can generate extra RM cells. These ''out-of-rate'' RM cells are not counted in the ACR of the source and are distinguished by having their cell loss priority (CLP) bit set, which means that the network will carry them only if there is plenty of bandwidth and can discard them if congested.
The out-of-rate RM cells generated by the source and switch are limited to 10 RM cells per second per VC. One use of out-of-rate RM cells is for BECN from the switches. Another use is for a source, whose ACR has been set to zero by the network, to periodically sense the state of the network. Out-of-rate RM cells are also used by destinations of VCs whose reverse direction ACR is either zero or not sufficient to return all RM cells received in the forward direction. Note that in-rate and out-of-rate distinction applies only to RM cells. All data cells in ABR should have CLP set to 0 and must always be within the rate allowed by the network.

Resource Management cells traveling from the source to the destination are called Forward RM (FRM) cells. The destination turns around these RM cells and sends them back to the source on the same VC. Such RM cells traveling from the destination to the source are called Backward RM (BRM) cells. Note that when there is bi-directional traffic, there are FRMs and BRMs in both directions on the Virtual Channel (VC). A direction bit (DIR) in the RM cell payload indicates whether it is an FRM or BRM.

## The ERICA Switch Scheme

The ERICA algorithm [SHIV] operates at each output port (or link) of a switch. The switch periodically monitors the load on each link and determines a load factor (z), the available ABR capacity, and the number of currently active virtual connections or VCs (N). A measurement or ''averaging'' interval is used for this purpose. These quantities are used to calculate the feedback which is indicated in RM cells. The feedback is given to the RM cells travelling in the reverse direction. Further, the switch gives at most one new feedback per source in any averaging interval. The key steps in ERICA are as follows:

**At the End of at Averaging Interval,** total ABR Capacity is computed as the difference between the link capacity and the bandwidth used by higher priority traffic. The Target ABR Capacity is then computed as a fraction (typically 0.9) of the total ABR capacity. The overload (Z) and the fair share (FS) are calculated as

```
z  ← ABR Input Rate / Target ABR Capacity
FS ← Target ABR Capacity / N
```

Where N is the number of active VCs. The maximum allocations of the previous current intervals are maintained as below:

```
MaxAllocPrevious ← MaxAllocCurrent
MaxAllocCurrent  ← FS
```

**When an FRM is received,** the Current Cell Rate

(CCR) in the RM cell is noted for the VC:

```
CCR[VC] ← CCR_in_RM_Cell
```

**When a BRM is received**
Feedback is calculated as follows and inserted in the ER field of the cell:

```
VCShare ← CCR[VC] / z
IF z > 1+ Δ
THEN ER ← Max (FairShare, VCShare)
ELSE ER ← Max (MaxAllocPrevious, VCShare)
MaxAllocCurrent ← Max(MaxAllocCurrent,ER)
IF (ER}>FairShare AND CCR[VC]<FairShare
     THEN ER ← FairShare
ER in RM Cell ← Min (ER in RM Cell, ER,
Target ABR
```

Details of the ERICA algorithm are available from [SHIV].

## The Switch Process Model

The OPNET process modeling methodology was used in the development of the switch process model that delivered basic capabilities of the core ATM switching fabric, ABR feedback control, buffer management and scheduling. The key steps of this modeling methodology include: definition of the system context, identifying interdependent modules, enumeration of events, selection of states of a process, construction of an event response table and construction of the finite state machine. The development of the OPNET Switch process model is described in the paragraphs below. A simple switch process receives cells on its input port. Cells are switched via the switching fabric to an output port based on its destination address. Cells may be enqueued at the output port and transmitted based on a scheduling algorithm. The functionality of a simple switch is illustrated in Figure 2.

Figure 2 A Simple ATM Switch

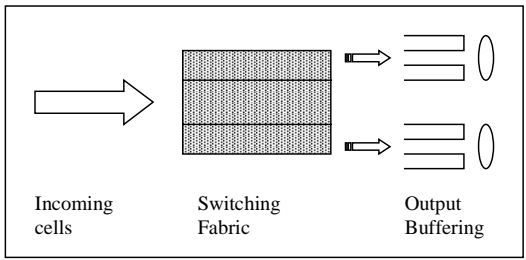

Incoming cells | Switching Fabric | Output Buffering

Logical events that can occur at the switch include a cell arrival, time to transmit as indicated by the scheduler and the end of fabric delay. Table 1 enumerates the events that can occur at the switch and the associated interrupt types.

Table 1. ABR switch events

| Logical Event | Event Description | Interrupt Type |
|---|---|---|
| Cell_Arrival | Arrival of an ATM cell at the switch | Stream |
| Time_to_send | Indication from the scheduler that it is time to transmit a cell | Self |
| End_of_fabric _delay | Indication that a cell has completed processing via the switch fabric | Self |

Table 2 outlines the actions taken when an event occurs within the switch. Each row of this table represents a combination of a state and an event and their associated conditions. Different actions performed for each combination and the resulting next state are listed.

Figure 3 illustrates the state machine obtained as a result of the event response table. Multiple state machines are used for modularity. The *Init* state is entered when the process receives a *begin simulation* interrupt. The switch buffer configuration specified by the user and the ABR attributes (VSVD mode, feedback scheme) are obtained and initialization functions are executed. The ATM models go through a configuration phase where topologies and interconnections are verified. The process then goes into the *wait* state where all subsequent processing of cells takes place. When a cell arrives, the process checks if this is application traffic arriving from the higher layer or if this is link traffic. Application traffic for an ABR connection goes through the source rules described in [JAIN]. Link traffic for ABR connections goes through destination rules [JAIN] if the VSVD mode is ON. Otherwise, it goes through the switching fabric. Once all cells have been through the switching fabric, they are processed by the output buffer management function where they may be enqueued or dropped. A scheduler sends out cells from the buffers onto the link based on the cell rate for the connection and the scheduling scheme.

## Summary

We have presented the ATM ABR traffic management model, and its implementation in OPNET. This model will replace the existing ATM model in OPNET. Simulation results on ATM ABR performance will be presented at the conference.

Table 2 ATM ABR event response table

| Current State | Logical Event | Condition | Action | Next State |
| --- | --- | --- | --- | --- |
| None | Begsim | None | None | Init |
| Init |  | None | Initialize | Config |
| Config | Cell_Arrival | Neighbor notification not complete | Queue cell | Config |
|  | Notify Complete | None | Process enqueued cells | Wait |
| Wait | Cell_arrival | Application traffic | Apply source rules before enqueue, schedule fabric delay | Wait |
|  |  | Link Traffic and VSVD_ON | Apply destination rules, apply source rules before enqueue, schedule fabric delay | Wait |
|  |  | Link_Traffic and VSVD_OFF | Schedule fabric delay | Wait |
|  | End_of_fabric_delay | Cell can be buffered | Enqueue cell, activate scheduler | Wait |
|  |  | Cell cannot be buffered | Destroy cell | Wait |
|  | Time_to_send | More cells waiting to be sent | Dequeue and send cell, re-activate scheduler | Wait |
|  |  | No more cells waiting | None | Wait |

Figure 3 OPNET ATM ABR state machine implementation

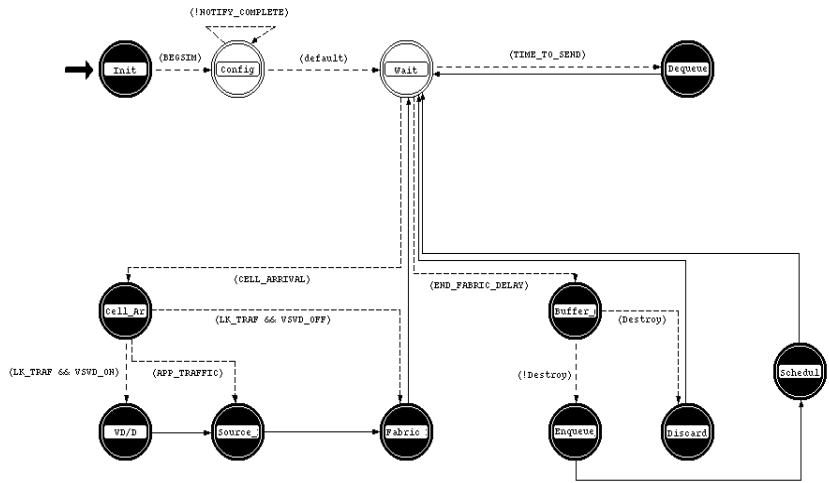

### References

[TM40] ATM Forum Traffic Management Specification, Version 4.0, ATM Forum, April 1996.

[JAIN] Raj Jain et. al, "Source Behavior for ATM ABR Traffic Management: An Explanation," IEEE Communications Magazine, November 1996.

[SHIV] Shivkumar Kalyanaraman, et. al, "The ERICA switch algorithm for ABR traffic management in ATM networks," submitted to IEEE/ACM Transactions on Networking.

[MIL31] OPNET Modeling Manual Vol 1, OPNET Version 3.5, MIL 3 Inc., 1997.

[MIL32] OPNET Models/Protocols Manual, OPNET Version 3.5, MIL 3 Inc., 1997.